\documentclass[10pt]{article}

\usepackage{hyperref}
\usepackage{amsmath, amsthm}
\usepackage{graphicx}

\title{Detecting Influenza Epidemics on Twitter}

\date{2019}

\author{
Katerina Katsani-Geronymaki\\
Computer Science Department, University of Crete
\and
Polyvios Pratikakis\\
Computer Science Department, University of Crete
}

\begin{document}

\maketitle

\begin{abstract}
This paper presents a predictive model for Influenza-Like-Illness,
based on Twitter traffic.  We gather data from Twitter based on a set
of keywords used in the Influenza wikipedia page, and perform feature
selection over all words used in 3 years worth of tweets, using real
ILI data from the Greek CDC.  We select a small set of words with high
correlation to the ILI score, and train a regression model to predict
the ILI score cases from the word features.  We deploy this model on a
streaming application and feed the resulting time-series to FluHMM, an
existing prediction model for the phases of the epidemic.  We find
that Twitter traffic offers a good source of information and can
generate early warnings compared to the existing sentinel protocol
using a set of associated physicians all over Greece.
\end{abstract}

\section{Introduction}

We aim to estimate the probability of a future Influenza or Influenza
Like Illness (ILI)\cite{flu}  epidemic using a stream of Twitter
posts\footnote{This work was performed in 2018 and 2019, and has now
been made obsolete by the COVID-19 pandemic, as any word-trained
predictive model does not apply, since the pandemic-related twitter
traffic dominates any other ILI-related traffic.}.
Monitoring and detecting an epidemic and its diffusion is a task of
great importance. It can help authorities and health experts plan
their responses, such as preparing specific Anti-Influenza drugs in
advance, or issue a warning towards vulnerable groups. As crucial as
monitoring is, it can be very challenging.  Information gathered from
the general public, such absences recorded in schools and work
environments, search engine queries about symptoms~\cite{ginsberg2009detecting}, or actual patient
reports from health practitioners, are often either inconclusive (one
could be missing work due to having the flu, or for another reason),
or delayed (processing and submitting such accumulated information
takes time).

Twitter~\cite{twitter} is a micro-blogging service that can
serve as a valuable information resource in this process of epidemic
surveillance. As shown by
Lampos et al.~\cite{lampos2010tracking,lampos2010flu},
Aramaki et al.~\cite{aramaki2011twitter},
Lee et al.~\cite{lee2013real},
and Achrekar et al.~\cite{achrekar2011predicting}, monitoring user
posts (tweets) offers a direct and real-time perspective to current
events, including Influenza symptoms or possible cases.  The
continuous growth of its community, along with the immediacy and the
convenience it offers users in expressing their thoughts and updating
their statuses, renders Twitter a timely, accurate and effective tool
for health research and surveillance.

In this paper we focus on the Greek-speaking twitter community and
influenza cases in Greece.  We integrate data from the twitter stream
into FluHMM~\cite{lytras2019fluhmm}, an open source R package for
seasonal influenza (ILI) sentinel surveillance that performs fitting
of a Bayesian HMM (Hidden Markov Model) to data.  Traditionally,
FluHMM uses patient reports of health practitioners and can calculate
the posterior probability of the next ILI epidemic phase. In order to
reduce processing times, we use FluHMM with input from Twitter (since
tweets can be accurate indications of the real situations), resulting
in a more time-efficient predicting model.  To do this, we use a large
corpus of offline twitter data to select a set of words which have a
high correlation with actual ILI recorded cases, using 3 years' worth
of historical ILI data from Greece.  We then train a predictive model
based on the selected set of word features and deploy it on a current
stream of Twitter traffic.

\section{Implementation} 

\subsection{Offline Twitter data}

Using twAwler~\cite{pratikakis2018twawler}, a lightweight twitter crawler that can
run on a single machine, we were able to obtain Greek-language tweets
containing keywords that are related to Influenza and express either
illness symptoms or describe the infection itself. Weekly aggregated
Twitter data from 2013 to 2015, along with the actual ILI patient
scores for the same time, were used to select the ILI keywords we will
be monitoring for our current approach.

\subsection{Feature Selection}

We used the greek Wikipedia\cite{grflu,grwho} page about Influenza to
obtain a large set of flu related terms, by eliminating stopwords in
the entire page, and keeping the rest. Then, a medical expert hand
picked the most relevant words out of those. We crawled all related
tweets and counted all instances of the selected terms for three
years.  To decide on the most relevant ILI terms, we calculate each
term's correlation to the actual ILI score presented by the country's
health organisation for those past years. The terms with the highest
Pearson’s correlation coefficient (Pearson's R), and therefore
the ones most strongly related to ILI, along with their coefficients,
are presented in Table~\ref{table1}.

\begin{table}[htp!]
\begin{center}

\begin{tabular}{ | c | c |}
\hline
 \textbf{Keyword} & \textbf{Pearson's R} \\
\hline
  \textbf{$\gamma\rho\iota\pi\eta$} & \textbf{0.4675661553} \\
\hline
 \textbf{$\gamma\rho\iota\pi\eta\varsigma$} & \textbf{0.3207126142} \\
\hline
 \textbf{$\kappa\rho\upsilon o \lambda o
 \gamma\eta\mu\alpha\tau\alpha$} & \textbf{0.3093922553} \\
\hline
 \textbf{$\kappa\rho\upsilon\omega\mu\alpha$} & \textbf{0.3719017249} \\
\hline
 \textbf{$\iota\omega\sigma\eta$} & \textbf{0.3671084547} \\
\hline
 \textbf{$\iota\omega\sigma\epsilon\iota\varsigma$} &
 \textbf{0.3933606774} \\
\hline
 \textbf{$\beta\eta\chi\alpha\varsigma$} & \textbf{0.3303753935} \\
\hline
 \textbf{$\beta\eta\chi\alpha$} & \textbf{0.3868373845} \\
\hline
 \textbf{$\beta\eta\chi\omega$} & \textbf{0.3269930131} \\
\hline \textbf{$\beta\eta\chi\epsilon\iota\varsigma$} &
\textbf{0.2838413272} \\
\hline
\end{tabular}
\caption{All keywords with the higher correlation to ILI, along with their Pearson's coefficients.}
\label{table1}
\end{center}
\end{table}

Clearly, a user tweeting one or more of the above words does not
necessarily indicates a person affected with the flu.  However, an
increase in overall occurence is correlated with ILI cases, over the
3-year period studied.  We trained a linear regression model with the
counters of those terms, and deploy the model on live data using a
streaming cralwer targeting the selected keywords.

\subsection{Collecting Tweets}
To apply the model trained on historic data and monitor the diffusion
of the disease, we collect a continuous stream of tweets created by
Greek users, which contain one or more ILI related terms. More
specifically, we expand our set of keywords to contain many versions of
each term, using combinations of capital letters, noun declinations
and verb conjugations, and accents.  This generates a streaming flow
of matching tweets, which however is not of huge volume, as most of
Twitter traffic is not illness-related.

This way we can obtain a list of public tweets (including their
actual content, along with the time they were posted, the user who
posted them and the location of the user) containing one or more of
our assigned keywords, the moment they are posted.  To evaluate the
application, we monitored the stream every day for one
month (30 days), resulting in an estimated number of ILI cases, by
applying the trained model to the daily twitter features.

\subsection{Hidden Markov Model for influenza surveillance}
FluHMM is an open source R package for seasonal influenza ILI sentinel
surveillance that performs fitting of a Bayesian HMM (Hidden Markov
Model) to data.  FluHMM constructs an inference model based
on the assumptions of a Markov process.
Traditionally, FluHMM uses surveillance data (number of patients
recorded) collected from a network of primary-care  physicians,
monitored by the Greek CDC, in order to partition the surveillance
period into five phases (pre ILI epidemic, epidemic growth, epidemic
plateau, epidemic decline and post-epidemic phase) and determine the
weekly posterior probability of each phase.  While accurate, FluHMM
results can be delayed, since processing patient reports can take
time, along with manually processing them, calculating ILI scores and
fitting weekly scores.

In our approach, instead of the number of actual patients, we produce
an estimated number of actual patients based on our aggregated Twitter
data, using a linear regression model trained on historical data.
Tweets correspond to real time events and there is no delay or
processing time. This can be crucial in cases of epidemic growth,
since authorities and health experts need to be alert in advance, and
the sooner we have results the better.

More specifically, after we have calculated the total of ILI-related
tweets per day  and performed linear regression to calculate a daily
estimated ILI score, we pass an array of ILI scores as input and
proceed to aggregate the scores per week. Then we start fitting the
model on a vector of  integers (one for each week), performing 5000
iterations. After the initial sampling is complete, the model proceeds
to check for initial convergence, and if it is not reached, it starts
updating in increments of 5000 iterations until convergence is
reached. Once convergence is reached, an object of class FluHMM is
created, containing the fitted model, which can then be plotted to
show each week's epidemic probability.

\begin{figure}
\centering
\includegraphics[width=\textwidth]{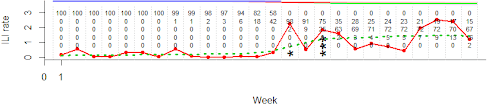}
\caption{A FluHMM fitted model for real ILI patients of June to December 2013}
\label{real13}
\end{figure}

\begin{figure}
\begin{center}
\includegraphics[width=\textwidth]{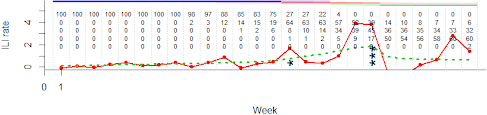}
\end{center}
\caption{A FluHMM fitted model for ILI related tweets of June to December 2013}
\label{fake13}
\end{figure}

\begin{figure}
\begin{center}
\includegraphics[width=\textwidth]{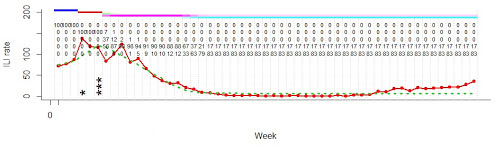}
\end{center}
\caption{A FluHMM fitted model for real ILI patients ofthe year 2015}
\label{real15}
\end{figure}

\begin{figure}
\begin{center}
\includegraphics[width=\textwidth]{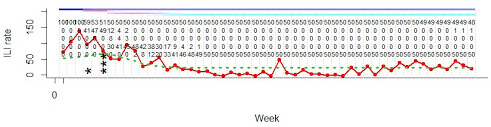}
\end{center}
\caption{A FluHMM fitted model for ILI related tweets of the year 2015}
\label{fake15}
\end{figure}

We show graphic representations of the results of fitting the second
semester of 2013 and the entirety of the year 2015, below.
Figure~\ref{real13} and Figure~\ref{real15} show the actual ILI
patient numbers for the years 2013 and 2015 respectively, compared to
the fitting of aggregated tweet count of ILI terms for the same
periods shown in Figure~\ref{fake13} and Figure~\ref{fake15}.
The posterior probabilities of the five epidemic phases per week
(pre-epidemic, epidemic growth,epidemic plateau, epidemic decline and
post-epidemic) are displayed with colored bars on the top of the plot
and also as a vertical stack of numbers.  Note that both the tweet
count and the actual patient count have a blue bar above the fourteen
first weeks displayed, which means both show that the first fourteen
weeks (June to mid September) are pre-epidemic. Additionally, each
column of numbers above every week indicates the percentage of each
phase probability.  

In the real ILI patient graph shown in Figure~\ref{real13}, weeks 1 to
13 have more than 80 percent pre epidemic probability, whereas weeks
14 to 17 have up to 98 percent chance of epidemic growth probability
(meaning people are starting to contract the flu) and weeks 18 to 26
(last two months of 2013) are divided between epidemic plateau, with
percentages from 63 percent to 72 percent, and epidemic growth (15 to
35 percent).

Similarly, in the twitter data graph shown in Figure~\ref{fake13},
the first 13 weeks display a 83 to 100 percent pre epidemic
probability, while weeks 14 to 17 have a 57 to 64 epidemic growth
probability, and weeks 18 to 26 are divided between epidemic plateau
and epidemic decline.

In the real ILI score of the entire year of 2015, shown in
Figure~\ref{real15}, week 4 marks the beginning of epidemic growth,
with 100\% posterior probability, while week 6 has a 98\% probability
of epidemic plateau.

In the graph for the ILI related tweets of 2015 shown in
Figure~\ref{fake15}, week 4 is also marked with epidemic plateau, and
week 6 beginning of the epidemic plateau phase, in accordance with the
real ILI patient reports of the same time.  Therefore, fitting the
real patient report numbers, and the estimated ILI scores calculated
from tweets, gives similar results to FluHMM, with an average
deviation of one week.  Note that the training data and evaluation
data overlap in this case, as the ILI score prediction model was
trained using the shown datasets.  To further evaluate its use, we
perform additional evaluation using a real-time stream of Tweets,
presented below.

\subsection{Live stream results}
\begin{figure}
\begin{center}
\includegraphics[width=\textwidth]{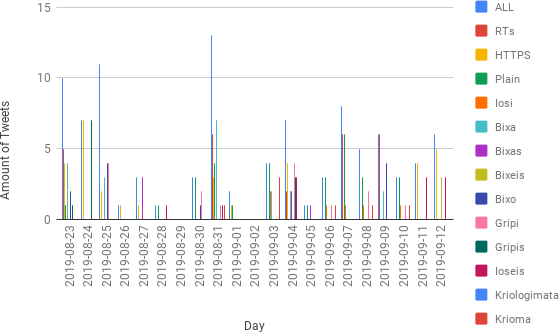}
\end{center}
\caption{Count of ILI related tweets Aug 23 to Sep 12 2019}
\label{chart19}
\end{figure}

\begin{figure}
\begin{center}
\includegraphics[width=0.5\textwidth]{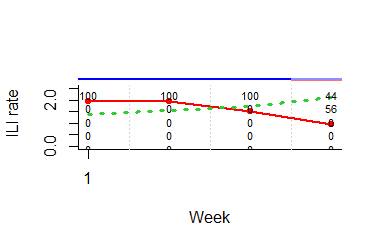}
\end{center}
\caption{FluHMM plot for estimated ILI score from Aug 23 to Sep 12 2019}
\label{flu19}
\end{figure}

We summarize and graphically plot our live data, to gain a collected
and better understanding of our findings, and also test how the
trained model can be applied in practice. That applies both to the raw
Twitter counters shown in Figure~\ref{chart19} and the FluHMM results,
shown in Figure~\ref{flu19}.

Figure \ref{chart19} shows a sample representation of our collected
testing data.  It shows the amount of Tweets posted containing any of
the selected ILI terms per day, from August 23 to September 12, 2019,
showing how many people tweeted ILI related posts on a span of three
weeks, as well as which specific term they mentioned.  The Figure
legend uses transliterations of the Greek words in the Latin alphabet.

Figure \ref{flu19} is a plot of our currently gathered data. Note that
the first two weeks (from August 23 to September 5) are 100 percent in
the pre epidemic phase, while the third week is starting to get in the
epidemic growth phase with a posterior probability of 56 percent.

\section{Conclusions}
Monitoring an Influenza epidemic is a vital task that can reduce
response times from authorities and health facilities, as well as help
vulnerable groups be alert and safe. Traditional ways of ILI
surveillance are often delayed due to processing times. Twitter is an
ever growing microblogging service that can aid in monitoring the
diffusion of an epidemic, as it allows users to share their current
status and situation, such as being sick with the flu, instantly.
We use a filtered Twitter stream based on selected word features and
find users that talk about Influenza daily, including one or more of
our selected ILI related terms in their posts. We then use weekly
counters to fit a Bayesian HMM (Hidden Markov Model) and calculate the
probability of an ILI epidemic for the following week, based on the
current week's situation. Though this model normally takes input from
health practitioners' reports on real ILI patients, when presented
with input from Twitter we can see very similar results, with
probability percentages having an average deviation of a week or less,
as shown in the 2013 and 2015 graphs. Therefore the approach of
substituting real ILI numbers with estimated ILI scores based on
counts of ILI related tweets, achieves a satisfactory sensitivity and
timeliness, thereby demonstrating its usefulness.

\bibliographystyle{plain}
\bibliography{main}

\end{document}